\documentclass[aps,preprint,nofootinbib,hyperref]{revtex4}%
\usepackage{amsfonts}
\usepackage{amsmath}
\usepackage{amssymb}
\usepackage{graphicx}%
\providecommand{\U}[1]{\protect\rule{.1in}{.1in}}
\providecommand{\U}[1]{\protect\rule{.1in}{.1in}}
\providecommand{\U}[1]{\protect\rule{.1in}{.1in}}
\providecommand{\U}[1]{\protect\rule{.1in}{.1in}}
\providecommand{\U}[1]{\protect\rule{.1in}{.1in}}
\providecommand{\U}[1]{\protect\rule{.1in}{.1in}}

\begin{document}
\preprint{ }
\title{Local Conformal Symmetry in Physics and Cosmology}
\author{Itzhak Bars}
\affiliation{Department of Physics and Astronomy, University of Southern California, Los
Angeles, CA, 90089-0484, USA,}
\author{Paul Steinhardt}
\affiliation{Physics Department, Princeton University, Princeton NJ08544, USA,}
\author{Neil Turok}
\affiliation{Perimeter Institute for Theoretical Physics, Waterloo, ON N2L 2Y5, Canada.}
\date{July 7, 2013}

\begin{abstract}
We show how to lift a generic non-scale invariant action in Einstein frame
into a locally conformally-invariant (or Weyl-invariant) theory and present a
new general form for Lagrangians consistent with Weyl symmetry. Advantages of
such a conformally invariant formulation of particle physics and gravity
include the possibility of constructing geodesically complete cosmologies. We
present a conformal-invariant version of the standard model coupled to
gravity, and show how Weyl symmetry may be used to obtain unprecedented
analytic control over its cosmological solutions. Within this new framework,
generic FRW cosmologies are geodesically complete through a series of big
crunch - big bang transitions. We discuss a new scenario of cosmic evolution
driven by the Higgs field in a \textquotedblleft minimal\textquotedblright%
\ conformal standard model, in which there is no new physics beyond the
standard model at low energies, and the current Higgs vacuum is metastable as
indicated by the latest LHC data.

\end{abstract}

\maketitle
\tableofcontents

\newpage

\section{Why Conformal Symmetry?}

Scale invariance is a well-known symmetry \cite{DiracETC} that has been
studied in many physical contexts. A strong physical motivation for
incorporating scale symmetry in fundamental physics \cite{Weyl}\cite{deser}%
\cite{englert} comes from low energy particle physics. Namely, the
\textit{classical} action of the standard model is already consistent with
scale symmetry if the Higgs mass term is dropped. This invites the idea, which
many have considered, that the mass term may emerge from the vacuum
expectation value of an additional scalar field $\phi\left(  x\right)  $ in a
fully scale invariant theory. Another striking hint of scale symmetry occurs
on cosmic scales: the (nearly) scale invariant spectrum of primordial
fluctuations, as measured by WMAP and the Planck satellite \cite{Planck,WMAP}.
This amazing simplicity seems to cry out for an explanation in terms of a
fundamental symmetry in nature, rather than as just the outcome of a scalar
field evolving along some particular potential given some particular initial condition.

In this paper, we consider the incorporation into fundamental physics of
\textit{local conformal symmetry} (or Weyl-symmetry): that is, classical local
scaling symmetry in an action that includes the standard model coupled to
gravity. A new result of our approach \cite{inflationBC}-\cite{BibBang-IB} is
that using conformal symmetry we are able to solve the classical FRW equations
across big crunch-big bang transitions, thus obtaining the\textit{ full set of
geodesically complete cosmological solutions} of our conformal standard model
given in Eq.(\ref{action1}). This follows from the new properties of the
standard theory whose couplings to gravity includes \textit{all patches of
field space} that are required for geodesic completeness of all cosmological
solutions for all times and any set of initial conditions\footnote{In this
paper, we use the term "geodesic completeness" to refer to two notions: (a)
geodesic continuation through all singularities separating patches of
spacetime; and (b) avoidance of unnatural initial conditions by requiring
infinite action for geodesics that reach arbitrarily far in the past. The two
notions are inequivalent since one is local in time and the other is a global
condition. Both properties are satisfied in our conformal standard model in
the form given in Eq.(\ref{action1}) as described in section (\ref{confCosm}%
).}.

The models we describe contain no mass scales -- no gravitational constant, no
mass for the Higgs field, no cosmological constant and no mass parameters for
the quarks, leptons or gauge bosons. All of these are prevented by the local
conformal symmetry combined with the SU$\left(  3\right)  \times$SU$\left(
2\right)  \times$U$\left(  1\right)  $ gauge symmetry of the standard model.
There is only one source of mass which follows from gauge fixing (in some
sense spontaneously breaking) the Weyl symmetry through a scalar field which
is a singlet under SU$\left(  3\right)  \times$SU$\left(  2\right)  \times
$U$\left(  1\right)  $ or a combination of both SU$\left(  3\right)  \times
$SU$\left(  2\right)  \times$U$\left(  1\right)  $ singlet and non-singlet
fields, depending on the choice of gauge.\ This source, which is associated
with the emergence of the Planck scale, drives the spontaneous breakdown of
the electroweak symmetry, which in turn generates the other known masses of
elementary constituents, with ratios of masses related to dimensionless
constants.

Our approach to conformal symmetry has new features with significant physical
consequences that were not considered before. While we shall work in $3+1$
dimensions throughout, there is a close connection with field theories and
gravity or supergravity in $4+2$ dimensions in the context of 2T-physics
\cite{2TphaseSpace}-\cite{ibsuper} as will be pointed out occasionally
throughout this paper. Stated directly in $3+1$ dimensions, some of the novel
features and their consequences are as follows:

\begin{itemize}
\item In section (\ref{general}) we provide the most general form of coupling
gravity or supergravity with any number of scalar fields, fermions and gauge
bosons, while maintaining local conformal symmetry. Only a specialized form of
our general formalism in section (\ref{general}) coincides with previously
known methods for local conformal symmetry. We find that in a gauge symmetric
theory, if there is only one physical scalar field (for example, the Higgs
field) \textit{after all gauge choices are exhausted}, then there is a unique
gauge-invariant and Weyl-invariant form for the Lagrangian representing the
geodesically complete coupling to gravity.

\item The minimal, realistic and locally conformal standard model coupled to
gravity includes the Higgs doublet $H\left(  x\right)  $ and an additional
SU$(3)\times$SU$(2)\times$U$(1)$ singlet $\phi\left(  x\right)  $. Taking into
account all of the local symmetry, there is a single physical scalar field
identifiable with the observed Higgs field. Therefore, according to the
statement above, this is a geodesically complete conformal standad model
provided it utilizes our unique coupling to gravity as shown in
Eq.(\ref{action1}) and as proven in section (\ref{generalGrav}). The relative
minus sign between $\phi$ and $H$ that appears in this action is obligatory
and is related to geodesic completeness as will be explained later. The unique
structure of local scale symmetry has led to full analytic control of
cosmological solutions and, in certain limits, enabled us to obtain the
complete set of homogeneous cosmological solutions for all possible initial
conditions of the relevant fields. This is discussed in section
(\ref{confCosm}).

\item In a particular gauge of the spontaneously broken local scale symmetry
(called $c$-gauge) we recover the usual renormalizable field theory of the
standard model in the flat space limit, including a mass scale for the Higgs.
In this gauge, low energy physics of our model coincides with the familiar
form of the standard model unless more fields are included beyond those
already observed, such as right handed neutrinos and dark matter which can be
accommodated consistently with conformal symmetry following our methods. Thus,
low energy physics at LHC scales is not sensitive to the conformal structures
suggested in our approach.

\item On the other hand, physics at cosmological scales can be quite sensitive
to the conformal structures discussed in this paper as becomes apparent in
certain gauges. In section (\ref{confCosm}) we outline some of the phenomena
that follow from our unique conformal coupling. In the minimal conformal
standard model of Eq.(\ref{action1}), guided by our analytic solutions
(supplemented, where necessary, with numerical methods), we propose a
\textit{cyclic conformal cosmology} driven only by the Higgs field with no
recourse to other scalar fields such as an inflaton. The Higgs-driven
cosmological scenario we propose is strongly motivated by the metastability of
the Higgs vacuum as implied by the renormalization-group flow of the minimal
standard model to large Higgs vev, with its parameters fixed by the LHC
measurements \cite{instabilityHiggs}.
\end{itemize}

The main discussion in this paper is at the classical field theory level, but
it is worth commenting briefly on the important question of quantum
corrections. There are two enormous hierarchies in the standard model coupled
to gravity: the weak scale is around $10^{-16}$ and the dark energy scale is
around $10^{-30}$ of the Planck scale. It is hard to understand how such tiny
numbers enter fundamental physics, and why quantum corrections would not spoil
these fine tunings.

In both cases, conformal symmetry has been suggested as a solution. We will
not explore this here in any detail, except to note that the stability of such
hierarchies in the perturbative standard model, which is what attracted
attention in the past for the Higgs field, should not be in jeopardy from
quantum corrections since dimensionless constants in a conformal theory are
logarithmically divergent as opposed to the quadratic divergence of a bare
Higgs mass term. To fully grasp how this can work at the quantum level for
fundamental scalars and in particular the Higgs, requires a better
understanding of how to perform regularization and renormalization consistent
with local conformal symmetry in $3+1$ dimensions. Elaborating on earlier
suggestions \cite{englert}\cite{bardeen}\cite{nicolai}, there has been recent
progress \cite{percacci} in developing a renormalization theory consistent
with Weyl symmetry. According to this new work, local conformal symmetry
remains as a valid symmetry at the quantum level as anticipated in
\cite{englert}. The local scale invariance survives quantization even though
there is a trace anomaly in the stress tensor of the matter sector of the
theory; this point that caused confusion in the past is clarified by noting
that the trace of the \textit{matter} stress tensor is \textit{distinct} from
the generator of local scale transformations that includes the additional
fields, such as $\phi,$ that implement and establish the local conformal
symmetry. In passing, we note that if $3+1$ dimensional conformal symmetry is
treated consistently within 2T-physics in $4+2$ dimensions, then scale
(dilation) symmetry is a part of a \textit{linearly realized} SO$\left(
4,2\right)  $ in the flat limit which is presumably not anomalous just as its
Lorentz subgroup SO$\left(  3,1\right)  $ in $3+1$ dimensions cannot be
anomalous. This quantum aspect of 2T-physics is under investigation directly
in $4+2$ dimensions and this is expected to shed additional light on these issues.

A plan of this paper is as follows. In Sec. II, we will review and expand on
the motivation for global and local symmetry and show how to recast (or
\textquotedblleft lift\textquotedblright) the standard model plus gravity into
a locally gauge invariant, Weyl- symmetric theory by adding new fields and, at
the same time, introducing compensating gauge symmetries, in such a way that
the theory is \textit{both Weyl invariant and geodesically complete}. The
gauge fixed version of this theory reverts back to the familiar minimal form
of the standard model at low energies wherever gravity is negligible. However,
in regions of spacetime where strong gravity effects are important, such as in
cosmology, especially close to the singularity, the geodesically complete Weyl
lifted version plays a crucial role, as described in Sec. IV.

In section III, following \cite{ibsuper} we present a general form for a
Lagrangian consistent with Weyl symmetry for any number of scalar fields. We
argue that, if the Weyl invariant theory is equivalent to having only a single
physical scalar field after gauge fixing (like the physical Higgs), then by
field redefinitions it is always possible to recast the theory into our
version given in Eq.(\ref{action1}) which includes all patches in field space
so as to insure geodesic completeness.

As another example of our general formalism, we lift the Bezrukov-Shaposhnikov
Higgs inflation model \cite{shoposh-bez} to a fully scale invariant model,
making it logically and physically consistent. Based on our formalism for
constructing general Weyl-invariant actions, we show that the model is not
unique. In fact, there are many single-field cosmological Higgs models with
the same properties, all consistent with conformal symmetry and producing the
same inflationary outcome. However, this inflationary scenario is not
geodesically complete, furthermore, in view of our complete set of solutions,
it is extremely unlikely.

In Sec. IV, we turn more generally to cosmology. We explain that a Weyl-lifted
theory can resolve the singularity and enable the geodesic completion of
cosmological spacetimes, while also indicating the presence of new phenomena.
Our previous work in \cite{inflationBC}-\cite{BibBang-IB} elaborated in some
detail on the properties of geodesically complete cosmological analytic
solutions in the context of solvable examples of what was supposed to be toy
models. However the same models now re-appear in our conformal standard model
and hence our previous solutions are now the full set of cosmological
solutions for all initial conditions of the familiar standard model. In
particular, we point out a model independent attractor mechanism discovered in
\cite{BCST1} that may help to determine the likely initial conditions of our
universe just after the big bang. The Weyl-invariant formulation also provides
a natural framework for incorporating cyclic cosmology \cite{cyclic1,cyclic2}
driven only by the Higgs field as introduced. In the case that the Higgs field
is metastable, the Higgs inflation model is inoperable, but the minimal
single-Higgs model is naturally compatible with the cyclic picture
\cite{lectures}. In general, the Weyl-invariant approach also hints at the
intriguing possibility that the minimal, electroweak Higgs may have played a
central role in cosmic evolution as explained in detail in
\cite{BST-HiggsCosmo}.

There is a deep connection between the ideas presented in this paper and
theories in $4$-space and $2$-time dimensions. Many of the ideas discussed
here for a global or local conformally symmetric theory emerged progressively
from developing the $4+2$ dimensional formalism since 1996 (for a recent
overview see \cite{2TphaseSpace}), and then for the standard model as given in
\cite{2Tsm}, for gravity in \cite{2Tgravity}, for SUSY in \cite{2Tsusy} and
supergravity in \cite{ibsuper}. The formulation given in \cite{ibsuper}, with
many scalars coupled to gravity and supergravity in $4+2$ and $3+1$ dimensions
was the precursor of the general formalism presented in section III, while the
simplest specific models were analyzed cosmologically in some detail in
\cite{inflationBC}-\cite{BibBang-IB}. There are some recent similar examples
\cite{feraraLindeKal} that overlap with our conformal symmetry vision in $3+1$
dimensions; these seem to be oblivious to our previous publications that
introduced at an earlier stage the crucial conformal structures with
restrictions on scalars in $3+1$ dimensions as predictions from 2T-physics
\cite{ibsuper}. The $4+2$ theory has more predictions of hidden symmetries and
dualities in $3+1$ dimensions which are mainly understood in classical and
quantum mechanical contexts \cite{2TphaseSpace} and are also partially
developed in field theory \cite{dualitiesFields}; these go well beyond
conformal symmetry in their implications for unification and the meaning of
spacetime and are bound to play an interesting role in future progress.

We emphasize that, despite the name, the physics content in the 2T-physics
formalism in $\left(  d+2\right)  $ dimensions is same as the physics content
in the standard 1T-physics formalism in $\left(  d- 1\right)  +1$ dimensions
except that 2T physics provides a holographic perspective and, due to a much
larger set of gauge symmetries, naturally makes predictions that are not
anticipated in 1T-physics. These additional gauge symmetries are in phase
space rather than position space, and can be realized only if the formalism is
developed with two times. Nevertheless, the gauge invariant sector of
2T-physics is equivalent to a causal 1-time spacetime without any ghosts.

In this paper, we will stick to $3+1$ dimensions, advocating a coherent
overall picture of a conformally invariant formulation of fundamental physics
and cosmology. However, we will occasionally remind the reader that these
outcomes naturally follow from $4+2$ dimensions with appropriate (but unusual)
gauge symmetries.

\section{Why Local Conformal Symmetry?}

In this section, we will describe the motivation for and construction of
simple theories with global and local scale invariance. An important
application is the lift of the standard model plus gravity into a Weyl-
invariant theory that is also geodesically complete.

\subsection{Global scale invariance \label{gscaleInv}}

We begin by examining a scale invariant extension of the standard model of
particles and forces. For example consider the usual standard model with all
the usual fields, including the doublet Higgs field $H\left(  x\right)  $
coupled to gauge bosons and fermions, but add also an SU$\left(  2\right)
\times$U$\left(  1\right)  $ singlet $\phi\left(  x\right)  $ (plus right
handed neutrinos and dark matter candidates), and take the following purely
quartic renormalizable potential involving only the minimal set of scalar
fields%
\begin{equation}
V\left(  H,\phi\right)  =\frac{\lambda}{4}\left(  H^{\dagger}H-\alpha^{2}%
\phi^{2}\right)  ^{2}+\frac{\lambda^{\prime}}{4}\phi^{4}. \label{potential}%
\end{equation}
This model, discussed at some length in section VI of \cite{2Tsm}, is the
minimal extension of the standard model that is fully scale invariant at the
classical level, globally. See also \cite{shaposh2}-\cite{jschwarz} that use a
similar field, and the different approach that also adds a Weyl vector field
\cite{cheng} leading to different physical consequences. The field
$\phi\left(  x\right)  $, which we call the \textquotedblleft
dilaton\textquotedblright, absorbs the scale transformations and is analogous
to the dilaton in string theory. In the current context, it has a number of
interesting features: Due to SU$\left(  3\right)  \times$SU$\left(  2\right)
\times$U$\left(  1\right)  $ gauge symmetry, the singlet $\phi$ is prevented
from coupling to all other fields of the Standard model - except for the
additional right handed singlet neutrinos or dark matter candidates. These
features of $\phi,$ that prevent it from interacting substantially with
standard visible matter except via the Higgs in Eq.(\ref{potential}), suggest
naturally that $\phi$ itself could be a candidate for dark matter \cite{2Tsm}.

The only parameters associated with $\phi,$ that are relevant for our
discussion, are $\alpha$ and $\lambda^{\prime}$. For positive $\lambda
^{\prime}$ the minimum of this potential occurs at $H^{\dagger}H=\alpha
^{2}\phi^{2}.$ Accordingly, the vacuum expectation value of the Higgs may
fluctuate throughout spacetime, depending on the dynamics of $\phi\left(
x\right)  ,$ without breaking the scale symmetry. However, if for some reason
(e.g. driven by quantum fluctuations or gravitational interactions) $\phi$
develops a vacuum expectation value $\phi_{0}$ which is constant in some
region of spacetime, then the Higgs is dominated by a constant vacuum
expectation value
\begin{equation}
H_{0}^{\dagger}H_{0}=\alpha^{2}\phi_{0}^{2}\equiv\frac{v^{2}}{2},
\label{v-value}%
\end{equation}
with $v$ fixed by observation to be approximately $246~GeV.$ The Higgs vacuum
$H_{0}$ provides the source of mass for all known elementary forms of matter,
quarks, leptons, and gauge bosons (while $\phi_{0}$ may be the source of
Majorana mass for neutrinos). The observation at the LHC of the Higgs
particle, which is just the small fluctuation on top of the vacuum value $v$,
has by now solidified the view that this is how nature works in our region of
the universe, at least up to the energy scales of the LHC.

Fig.~1 illustrates how the Higgs field slowly relaxes to the spontaneously
broken symmetry vacuum described by Eq.~(\ref{v-value}) beginning from large
oscillations shortly after the big bang. Here, as is the case throughout the
paper, the solutions are in the limit of negligible gauge and top quark mass
coupling so that the Higgs evolution is described by classical equations of
motion. This time-dependent behavior of the Higgs is driven by the evolution
of the field $\phi$, as anticipated in \cite{2Tsm2} for this simple model. In
fact, it is realized as the generic solution for all homogeneous cosmological
solutions if the Higgs vacuum is stable.

There are other interesting theoretical structures worth noting about this
scale invariant setup. Instead of supersymmetry, conformal symmetry may
explain the stability of the hierarchy between the low mass scale of $246~GeV$
versus the Planck scale of $10^{19}$ GeV, as suggested in \cite{2Tsm2}. The
conformal protection of the hierarchy is not as clear-cut as SUSY's protection
and requires better understanding of regularization and renormalization
techniques consistent with conformal symmetry as outlined in the introduction.

What about the dilaton (fluctuations in $\phi$) that emerges in the broken
scale invariance scenario above? At least from the perspective of only the
standard model, the dilaton is a massless Goldstone boson due to the
spontaneous breaking of the global scale invariance. As discussed in some
detail in section VI of \cite{2Tsm}, the original doublet field $H$ is the
only field coupled to known matter, while $\phi$ is decoupled. However, after
the spontaneous breaking, $H$ must be rewritten as a mixture of the mass
eigenstates of the model, which include the observed massive Higgs particle
and the massless dilaton (or maybe low-mass dilaton if the scale symmetry is
broken by some source). The mixing strength is controlled by the dimensionless
parameter
\begin{equation}
\alpha=(246~GeV)/\left(  \sqrt{2}\phi_{0}\right)  , \label{alpha}%
\end{equation}
that appears in Eq.(\ref{v-value}). Therefore, through this mixing, the
(pseudo-) Goldstone dilaton must couple to all matter, just like the Higgs
does, namely with a coupling given by the mixing angle $\left(  \mathrm{sin}%
(\theta)=\alpha/(1+\alpha^{2})^{1/2}\right)  $ times the mass of the particle
divided by $v$. The largest coupling is to the top quark. If $\phi_{0}$ is
much larger than $246$ GeV$,$ it is possible for the dilaton to hide from
accelerator experiments due to the weak coupling of $\alpha$ in
Eq.(\ref{alpha}).

Such a (pseudo) Goldstone boson has many other observable consequences,
including a long range force that competes with gravity and contributions to
quantum effects as a virtual particle in Feynman loop diagrams. If such an
additional massless (or low mass scalar) is observed in experiments, it would
be strong evidence in favor of the scale invariant scenario. However, if it is
not observed, this could be interpreted merely as setting a lower limit on the
scale $\phi_{0}.$

Eventually it must be understood what sets the scale $\phi_{0}$. The model as
presented above has no mechanism at the classical level to set the scale
$\phi_{0}$; its equations of motion are self consistent at the minimum of the
potential, $H^{\dagger}_{0}H_{0}=\alpha^{2}\phi^{2}_{0},$ only if
$\lambda^{\prime}=0$ \cite{2Tsm}; then $\phi_{0}$ remains undetermined due to
a flat direction in the remaining term in the potential (\ref{potential}). The
value for $\phi_{0}$ would then be obtained phenomenologically from
experiment, without a theoretical explanation. Quantum corrections, such as
those discussed in \cite{Coleman}, may alleviate this problem by removing the
flat direction. However, if the quantum corrections are small, there is the
danger that $\phi_{0}$ would be so low that it is not possible to obtain the
small value of $\alpha$ in Eq.(\ref{alpha}) required to protect the dilaton
from current experimental limits.

\subsection{Local scale invariance \label{lscaleInv}}

Since at present there is no sign of the dilaton in low energy physics,
suppose it does not exist at all as a degree of freedom. Is this incompatible
with the idea that conformal symmetry underlies fundamental physics? Not at
all, because any possible phenomenological problems associated with a
Goldstone boson fluctuation of $\phi$ can be overcome if the scaling symmetry
is a \textit{local} gauge symmetry, known as the Weyl symmetry. Then the
massless fluctuations of the dilaton can be eliminated by fixing a unitary
gauge\footnote{The interesting features of the dilaton of the previous
section, including massive fluctuations, could emerge from one more
SU$(2)\times$U$\left(  1\right)  $ scalar field as part of the general theory
that we will discuss in section (\ref{generalGrav}). But for simplicity in
this section we will first concentrate on the minimal model that contains only
the confirmed observed degrees of freedom up to now, thus allowing the
fluctuations of a single $\phi$ to be eliminated by a gauge symmetry in this
minimal model.}.

The standard model decoupled from gravity has no local scale symmetry that
could remove a Goldstone dilaton. But such a gauge symmetry can in fact be
successfully incorporated as part of the standard model provided it is coupled
to gravity in the right way. Coupling the standard model to Einstein gravity
in the conventional way makes no sense because the dimensionful Newton
constant explicitly breaks scale invariance. If scale invariance is already
broken in one sector of the theory, then there is no rationale for requiring
that it be a good symmetry in another part of the theory. At best, it would
occur as an accidental symmetry of low energy physics and only when gravity is
negligible. This is not the scenario we have in mind; we argue for a fully
scale invariant approach to all physics, a natural outcome of the larger gauge
symmetries in $4+2$ dimensions, as formulated in 2T-physics. Happily, the idea
of an underlying $4+2$ dimensions with appropriate extra gauge symmetry fits
all known physics in $3+1$ dimensions, from dynamics of particles and field
theory \cite{2TphaseSpace}-\cite{2Tsusy} all the way to supergravity
\cite{ibsuper}. So, consistency with this larger underlying structure is well motivated.

In fact, there is a locally scale invariant field theory in $3+1$ dimensions,
compatible with 2T-physics in $4+2$ dimensions, that couples the standard
model and gravity with no dimensionful constants. We do not mean conformal
gravity which has ghost problems, but rather the non-minimal conformal
coupling of the curvature $R\left(  g\right)  $ to scalar fields \cite{deser}
which is invariant under Weyl transformations as a gauge symmetry. In the next
section we will discuss a generalized Weyl invariant coupling with many scalar
fields also predicted by 2T-physics, but in this section we begin with the
well known method of conformally coupled scalars, as follows \cite{deser}%
\cite{englert},%
\begin{equation}
\frac{1}{12}\phi^{2}R\left(  g\right)  +\frac{1}{2}g^{\mu\nu}\partial_{\mu
}\phi\partial_{\nu}\phi. \label{confScalar}%
\end{equation}
These two terms form an invariant unit (up to a total derivative) under
\textit{local} scale transformations, $g_{\mu\nu}\rightarrow\Omega^{-2}%
g_{\mu\nu},\;\phi\rightarrow\Omega\phi$, with a local parameter $\Omega\left(
x\right)  .$ When there are more scalar fields, the most general way of
achieving local conformal symmetry is discussed in section (\ref{GenWeyl}).
However, when there is only one additional scalar field beyond $\phi,$ which
is in a single representation of a Yang-Mills gauge group, there is a unique
way to also achieve geodesic completeness as explained in more detail in
sections (\ref{generalGrav},\ref{confCosm}). Hence, \textit{for geodesic
completeness} we require that not only the field $\phi,$ but also the doublet
Higgs field be a set of conformally coupled scalars consistent with SU$\left(
2\right)  \times$U$\left(  1\right)  $. Namely using the unit $\frac{1}%
{6}\left(  H^{\dagger}H\right)  R\left(  g\right)  +g^{\mu\nu}D_{\mu
}H^{\dagger}D_{\nu}H$ which is also locally invariant (up to a total
derivative), we can lift the globally scale invariant standard model of the
last section into a locally invariant one, while also being coupled to gravity
in a gedesically complete theory. All other terms present in the usual
standard model, namely all fermion, gauge boson and Yukawa terms, when
minimally coupled to gravity are already automatically invariant under the
local Weyl symmetry.

Hence a \textit{Weyl invariant} action $S=\int d^{4}x\mathcal{L}\left(
x\right)  $ that describes the coupling of gravity and the standard model is
given by
\begin{equation}
\mathcal{L}\left(  x\right)  =\sqrt{-g}\left[
\begin{array}
[c]{c}%
\frac{1}{12}\left(  \phi^{2}-2H^{\dagger}H\right)  R\left(  g\right)  \\
+g^{\mu\nu}\left(  \frac{1}{2}\partial_{\mu}\phi\partial_{\nu}\phi-D_{\mu
}H^{\dagger}D_{\nu}H\right)  \\
-\left(  \frac{\lambda}{4}\left(  H^{\dagger}H-\alpha^{2}\phi^{2}\right)
^{2}+\frac{\lambda^{\prime}}{4}\phi^{4}\right)  \\
+L_{\text{SM}}\left(
\begin{array}
[c]{c}%
\text{{\small quarks, leptons , gauge bosons,}}\\
\text{{\small Yukawa~couplings to~}}{\small H\&\phi,~}%
\text{{\small dark~matter.}}%
\end{array}
\right)
\end{array}
\right]  \label{action1}%
\end{equation}
The only Yukawa couplings of $\phi$ allowed by SU$\left(  3\right)  \times
$SU$\left(  2\right)  \times$U$\left(  1\right)  $ are to the right handed
neutrinos for which it becomes a source of mass. Note the relative minus sign
between $\phi$ and $H$ terms, which is required, as explained below. Here
$L_{\text{SM}}$ is the well known standard model Lagrangian minimally coupled
to gravity, except for the Higgs kinetic and potential terms, which are now
modified and explicitly written out in the first three lines. This action is
invariant under Weyl rescaling with an arbitrary local parameter
$\Omega\left(  x\right)  $ as follows%
\begin{equation}%
\begin{array}
[c]{c}%
g_{\mu\nu}\rightarrow\Omega^{-2}g_{\mu\nu},~\phi\rightarrow\Omega
\phi,\;H\rightarrow\Omega H,\;\\
\psi_{q,l}\rightarrow\Omega^{3/2}\psi_{q,l},\;A_{\mu}^{\gamma,W,Z,g}%
\rightarrow\Omega^{0}A_{\mu}^{\gamma,W,Z,g},
\end{array}
\label{WeylRescaling}%
\end{equation}
where $\psi_{q,l}$ are the fermionic fields for quarks or leptons, and
$A_{\mu}^{\gamma,W,Z,g}$ are the gauge fields for the photon, gluons, $W^{\pm
}$ and $Z.$ Note that the gauge bosons do not change under the Weyl rescaling.

We note that the Lagrangian (\ref{action1}) is the one obtained in the second
reference in \cite{2Tgravity} from the $4+2$ version of the standard model
\cite{2Tsm}\cite{2Tsm2} coupled to 2T- gravity \cite{2Tgravity}, by a method
of gauge fixing and solving some kinematical equations associated with
constraints related to the underlying gauge symmetries. In that approach we
learned that the Weyl symmetry is not an option in $3+1$ dimensions, it is a
prediction of 2T-physics: the $4+2$ dimensional theory is not Weyl invariant,
but yet the local Weyl symmetry in $3+1$ dimensions emerges as a remnant gauge
symmetry associated with the general coordinate transformations as they act in
the extra $1+1$ dimensions. So the Weyl symmetry is a required symmetry in
$3+1$ dimensions as predicted in the $4+2$ dimensional approach; this symmetry
carries information and imposes properties related to the extra $1+1$ space
and time dimensions \cite{2TphaseSpace}.

The Weyl gauge symmetry of the action in (\ref{action1}) does not allow any
dimensionful constants: no mass term in the Higgs potential, no
Einstein-Hilbert term with its dimensionful Newton constant, or any other mass
terms. This is a very appealing starting point because it leads to the
emergence of all dimensionful parameters from a single source. That source is
the field $\phi\left(  x\right)  $ that motivated this discussion in the
previous section, and the only scale is then generated by gauge fixing $\phi$
to a constant for all spacetime
\begin{equation}
\phi\left(  x\right)  \rightarrow\phi_{0}. \label{c-gauge}%
\end{equation}
In the gauge fixed version of the Lagrangian where $\phi\left(  x\right)
\rightarrow\phi_{0},$ we can express the physically important dimensionful
parameters, namely, the Newton constant $G$, the cosmological constant
$\Lambda$ associated with dark energy, and the electro-weak scale $v$, in
terms of $\phi_{0}$:
\begin{equation}
\frac{1}{16\pi G}=\frac{\phi_{0}^{2}}{12},\;\frac{\Lambda}{16\pi G}%
=\lambda^{\prime}\phi_{0}^{4},\;H_{0}^{\dagger}H_{0}=\alpha^{2}\phi_{0}%
^{2}\equiv\frac{v^{2}}{2}. \label{dimensionfulConstants}%
\end{equation}

Since the field $\phi\left(  x\right)  $ in this gauge ceases to be a degree
of freedom altogether, the massless dilaton is absent and the potential
problem with global scaling symmetry is avoided. Nevertheless, there still is
an underlying hidden conformal symmetry for the full theory.

We can explain now why it is necessary to have the opposite signs for $\phi$
and $H$ in the first two lines of the action in Eq.(\ref{action1}). The
positive sign for $\phi$ is necessary in order to obtain a positive
gravitational constant in (\ref{dimensionfulConstants}). However, conformal
symmetry then forces the kinetic term for $\phi$ to have the wrong sign, so
$\phi$ is a ghost. This can be seen to be a gauge artifact, though, since in a
unitary gauge the ghost $\phi$ is fixed to a constant or expressed in terms of
other degrees of freedom. This also explains why $H$ must have the opposite
sign, since otherwise $H$ would be a real ghost. This relative sign has
important consequences in cosmology as follows.

In flat space $R\left(  g\right)  \rightarrow0,$ where experiments such as
those at the LHC are conducted, the Lagrangian above becomes precisely the
usual standard model, including the familiar tachyonic mass term for the
Higgs. Furthermore, in weak gravitational fields, at low energies, the
gravitational effect of the Higgs field coupling to the curvature, $\frac
{1}{12}\left(  \phi_{0}^{2}-2H^{\dagger}H\right)  R\left(  g\right)  ,$ is
ignorable since $H$ (order of $v\approx246$ GeV) is tiny compared to the
Planck scale ($\phi_{0}\approx$10$^{19}$ GeV). Actually, the gravitational
constant measured at low energies is corrected by the electroweak scale,
namely $\left(  16\pi G\right)  ^{-1}=$ $\left(  \phi_{0}^{2}- v^{2}\right)
/12$ instead of (\ref{dimensionfulConstants}), but in practice this is a
negligible correction since $v^{2}\ll\phi_{0}^{2}.$ So, at low energies there
is no discernible difference between our Weyl-lifted theory and the usual
standard model. The practically isolated standard model appears as a
renormalizable theory decoupled from gravity.

However, in the cosmological context, the conformally coupled $H$ can and will
be large at some stages in the evolution of the universe. Working out the
dynamics of cosmological evolution, it turns out that, for generic initial
conditions, $H^{\dagger}H$ typically grows to large scales over cosmological
times, hitting $\left(  \phi_{0}^{2}-2H^{\dagger}H\right)  \rightarrow0$ in
the vicinity of a big bang or big crunch singularity. This behavior plays an
essential role in cosmological evolution, as well as in the resolution of the
singularity via geodesic completeness. In our conformal theory, the standard
model is not isolated from gravity, and we posit that the Higgs field can play
a bigger role in nature than originally anticipated.

We may take the Higgs doublet in the unitary gauge of the SU$\left(  2\right)
\times$U$\left(  1\right)  $ gauge symmetry
\begin{equation}
H\left(  x\right)  =\left(
\begin{array}
[c]{c}%
0\\
\frac{1}{\sqrt{2}}s\left(  x\right)
\end{array}
\right)  . \label{unitary}%
\end{equation}
The field $s\left(  x\right)  =v+\delta h\left(  x\right)  $, where $\delta
h\left(  x\right)  $ is the Higgs field fluctuation on top of the electroweak
vacuum, is then identified with the generic field $s\left(  x\right)  $ that
appeared in our cosmological papers \cite{inflationBC}-\cite{BibBang-IB}. The
action for $s(x)$ has the exact same form in those papers as here.

Thus, our previous analytical cosmological solutions can now be applied to
investigate the cosmological properties of the Higgs field. When the parameter
$\alpha$ vanishes, our previous work provides the full set of geodesically
complete analytic cosmological solutions for all initial conditions of the
relevant fields in the standard model coupled to gravity. The non-zero
$\alpha$ is then easily taken into account with numerical methods. This much
analytic control over cosmological properties of a realistic theory is
unprecedented. This became a very valuable tool that led us into a new
cosmological scenario driven only by the Higgs field, as outlined in section
(\ref{confCosm}) and completed in \cite{BST-HiggsCosmo}.

\section{General Weyl Symmetric Theory \label{general}}

The gauge symmetries derived from 2T-gravity and 2T-supergravity in $4+2$
dimensions lead to the general Weyl invariant coupling described below for any
number of scalar fields in $3+1$ dimensions \cite{ibsuper}. This generalizes
the possible forms of conformally coupled scalar theories beyond those
encompassed by Eq.(\ref{confScalar}) and allows for richer possibilities for
model building consistent with local scale invariance. We ignore the spinors
and gauge bosons whose minimal couplings to gravity are already automatically
Weyl symmetric.

\subsection{Gravity \label{generalGrav}}

We begin with gravity without supersymmetry. In the next subsection we will
indicate the additional constraints that emerge in supergravity. We assume any
number of \textit{real} scalar fields $\phi^{i}\left(  x\right)  $ labeled by
the index $i.$ If there are complex fields, we can extract their real and
imaginary parts and treat those as part of the $\phi^{i}.$ We introduce a Weyl
factor $U\left(  \phi^{i}\right)  ,$ a sigma-model-type metric in field space
$G_{ij}\left(  \phi^{i}\right)  $ and a potential $V\left(  \phi^{i}\right)
.$ The general Lagrangian takes the form%

\begin{equation}
\mathcal{L}=\sqrt{-g}\left(  \frac{1}{12}U\left(  \phi^{i}\right)  R\left(
g\right)  -\frac{1}{2}G_{ij}\left(  \phi^{i}\right)  g^{\mu\nu}\partial_{\mu
}\phi^{i}\partial_{\nu}\phi^{j}-V\left(  \phi^{i}\right)  \right)  .
\label{GenWeyl}%
\end{equation}
The results given in \cite{ibsuper} are the following constraints on these functions.

\begin{itemize}
\item $U\left(  \phi^{i}\right)  $ must be homogeneous of degree two,
$U\left(  t\phi^{i}\right)  =t^{2}U\left(  \phi^{i}\right)  ;$ and $V\left(
\phi^{i}\right)  $ must be homogeneous of degree four$,V\left(  t\phi
^{i}\right)  =t^{4}V\left(  \phi^{i}\right)  ;$ and $G_{ij}\left(  \phi
^{i}\right)  $ must be homogeneous of degree zero$,G_{ij}\left(  t\phi
^{i}\right)  =G_{ij}\left(  \phi^{i}\right)  .$

\item The following differential constraints must also be satisfied. These may
be interpreted as homothety conditions on the geometry in field space%
\begin{equation}
\partial_{i}U=-2G_{ij}\phi^{j},\;~\phi^{i}\partial_{i}U=2U,\;G_{ij}\phi
^{i}\phi^{j}=-U. \label{homothety}%
\end{equation}
The second and third equations follow from the first one and the homogeneity requirements.

\item Physics requirements also include that $G_{ij}$ cannot have more than
one negative eigenvalue because the local scale symmetry is just enough to
remove only one negative norm ghost. However, if more gauge symmetry that can
remove more ghosts is incorporated, then the number of negative eigenvalues
can increase accordingly. The gauged R-symmetry in supergravity (which is
automatic in the $4+2$ approach) is such an example.
\end{itemize}

In \cite{ibsuper} these rules emerged from gauge symmetries in 2T- gravity.
Since Weyl symmetry is an automatic outcome from $4+2$ dimensions, we can
check that these same requirements follow directly in $3+1$ dimensions by
imposing Weyl symmetry on the general form in Eq.(\ref{GenWeyl}).

As an example, it is easy to check that all these conditions are automatically
satisfied by the action given in Eq.(\ref{action1}), with one $\phi$ and four
real fields in the doublet $H.$ Using the symbol $s,$ as $s^{2}\equiv
2H^{\dagger}H,$ we write it in the form
\begin{equation}
U=\phi^{2}-s^{2},~V=\phi^{4}f\left(  s/\phi\right)  ,\;G_{ij}=\eta_{ij},
\label{standardForm}%
\end{equation}
where $\eta_{ij}$ is a flat Minkowski metric in the 5-dimensional field space
with a single negative eigenvalue - this reduces to a 2-dimensional $\eta
_{ij}$ in the unitary gauge of Eq.(\ref{unitary}) since indeed there is only a
single physical field $s$. In our work \cite{inflationBC}-\cite{BibBang-IB}
generally we let the potential $f\left(  z\right)  $ to be an arbitrary
function of its argument $z=s/\phi=\sqrt{2H^{\dagger}H}/\phi,$ as allowed by
the Weyl symmetry conditions above. In Eq.(1) we have a purely quartic
renormalizable potential, $V=\frac{\lambda}{4}\left(  s^{2}-\alpha^{2}\phi
^{2}\right)  ^{2}+\frac{\lambda^{\prime}}{4}\phi^{4}$. When quantum
corrections are included, $f\left(  z\right)  $ contains logarithmic
corrections, but by re-examining the underlying symmetry, the effective
quantum potential can again be rewritten in the form $\phi^{4}f\left(
s/\phi\right)  ,$ consistent with the local conformal symmetry. We use the
quantum corrected potential in our discussion of Higgs cosmology.

Next we give the general solution of the requirements above in a convenient
parametrization for $n+1$ fields labelled with $i=0,1,2,\cdots,n,$ namely,
$\phi^{i}=\left(  \phi,s^{I}\right)  ,$ where $\phi^{0}\equiv\phi$ is
distinguished, while $s^{I}$ with $I=1,2,\cdots,n$ are all the other scalar
fields. Then the general solution to the conditions above for Weyl gauge
symmetry takes the following form
\begin{align}
U\left(  \phi,s^{I}\right)   &  =\phi^{2}u\left(  z\right)  ,\text{ with any
}u\left(  z^{I}\right)  ,\text{ }\label{sol-u}\\
V\left(  \phi,s^{I}\right)   &  =\phi^{4}f\left(  z\right)  ,\;\text{with any
}f\left(  z^{I}\right)  ,\\
G_{IJ}\left(  z^{I}\right)   &  =\text{ any non-singular }n\times n\text{
metric,}\label{solG}\\
G_{0I}\left(  z\right)   &  =G_{I0}\left(  z\right)  =-\frac{1}{2}\left(
\frac{\partial u}{\partial z^{I}}+2G_{IK}z^{K}\right)  ,\\
G_{00}\left(  z\right)   &  =-u+z^{I}\frac{\partial u}{\partial z^{I}}%
+z^{I}z^{J}G_{IJ}, \label{sol-G00}%
\end{align}
where the ratio $z^{I}\equiv s^{I}/\phi$ is gauge invariant. Thus, after using
the chain rule, $\partial/\partial z^{I}=\phi~\partial/\partial s^{I},$ the
general Weyl invariant action becomes
\begin{equation}
\mathcal{L}=\sqrt{-g}\left(
\begin{array}
[c]{c}%
\frac{1}{12}U\left(  \phi,s\right)  R\left(  g\right)  -V\left(  \phi,s\right)
\\
+\frac{1}{2}\left(  U-s^{I}\frac{\partial U}{\partial s^{I}}-s^{K}s^{L}%
G_{KL}\right)  \left(  \partial^{\mu}\ln\phi\right)  \left(  \partial_{\mu}%
\ln\phi\right) \\
-\frac{1}{2}G_{IJ}\left(  \partial^{\mu}s^{I}\partial_{\mu}s^{J}\right)
+\left(  2G_{IJ}s^{J}+\frac{\partial U}{\partial s^{I}}\right)  \partial^{\mu
}s^{I}\partial_{\mu}\ln\phi
\end{array}
\right)  \label{WeylGrav}%
\end{equation}
with the $U,V,G_{IJ}$ in Eqs.(\ref{sol-u}-\ref{solG}). It should be noted that
$G_{IJ}\left(  z^{I}\right)  ,u\left(  z^{I}\right)  ,f\left(  z^{I}\right)  $
are not determined by Weyl symmetry alone. Various other symmetries in a given
model could restrict them. Any choice consistent with additional symmetries is
permitted in the construction of physical models. For example, the model in
Eq.(\ref{action1}) has the SU$\left(  3\right)  \times$SU$\left(  2\right)
\times$U$\left(  1\right)  $ symmetry. In particular, local superconformal
symmetry gives more severe restrictions by relating $G_{ij}$ and $U$ from the
beginning, as discussed below.

As the number of scalar fields increases, the restrictions imposed by Weyl
symmetry become less severe. For example, an additional SU$\left(  3\right)
\times$SU$\left(  2\right)  \times$U$\left(  1\right)  $ gauge singlet field
beyond $\phi$ that would be needed to reproduce the phenomenology of the
dilaton-like singlet discussed in section (\ref{gscaleInv}) can be included
with slightly more freedom on its coupling parameters. Keeping such
possibilities in mind for more general phenomenological considerations, we the
define the minimal model to include only the standard Higgs doublet and the
singlet $\phi$ as in the previous section.

Hence, for clarity we will write out the general Weyl invariant Lagrangian for
the case of only the Higgs doublet field $H$ plus $\phi.$ This generalizes
Eq.(\ref{action1}), but we suppress the other fields in our discussion.
Furthermore, we will work directly with the gauge fixed version of the Higgs
field in Eq.(\ref{unitary}), so the Higgs doublet is reduced to a single field
$s\left(  x\right)  $ as in (\ref{unitary}). In that case, from
Eqs.(\ref{sol-u}-\ref{sol-G00}) we obtain
\begin{equation}
\mathcal{L}\left(  x\right)  =\sqrt{-g}\left[
\begin{array}
[c]{c}%
\frac{1}{12}\phi^{2}u\left(  s/\phi\right)  R\left(  g\right)  -\phi
^{4}f\left(  s/\phi\right) \\
+g^{\mu\nu}\left(
\begin{array}
[c]{c}%
\frac{1}{2}\left(  u-\frac{s}{\phi}u^{\prime}-\frac{s^{2}}{\phi^{2}}\bar
{G}\right)  \partial_{\mu}\phi\partial_{\nu}\phi\\
-\frac{1}{2}\bar{G}\partial_{\mu}s\partial_{\nu}s+\left(  u^{\prime}+2\frac
{s}{\phi}\bar{G}\right)  \partial_{\mu}\phi\partial_{\nu}s
\end{array}
\right)
\end{array}
\right]  \label{action2}%
\end{equation}
where $G_{IJ}$ reduces to $G_{11}(s)\equiv\bar{G}(s)$ for the single field.

Generally, $\bar{G}\left(  s/\phi\right)  ,u\left(  s/\phi\right)  ,f\left(
s/\phi\right)  $ in Eq.(\ref{action2}) are three arbitrary functions of the
Higgs field, $z=s/\phi=\sqrt{2HH}/\phi,$ which may be used for model building.
As an example, if we take $U\left(  \phi,s\right)  =\phi^{2}+\xi s^{2},$ and
$\bar{G}=1,$ we obtain a relatively simple kinetic term with an arbitrary
potential%
\begin{equation}
\mathcal{L}\left(  x\right)  =\sqrt{-g}\left[
\begin{array}
[c]{c}%
\frac{1}{12}\left(  \phi^{2}+\xi s^{2}\right)  R\left(  g\right)  -\phi
^{4}f\left(  s/\phi\right) \\
+\left(
\begin{array}
[c]{c}%
\frac{1}{2}\left(  1-\frac{s^{2}}{\phi^{2}}\left(  1+\xi\right)  \right)
\partial_{\mu}\phi\partial^{\mu}\phi\\
-\frac{1}{2}\partial_{\mu}s\partial_{\nu}s+2\frac{s}{\phi}\left(
1+\xi\right)  \partial_{\mu}\phi\partial^{\mu}s
\end{array}
\right)
\end{array}
\right]  \label{xi}%
\end{equation}
This Lagrangian is Weyl invariant for any value of the constant parameter
$\xi,$ but we see now a simple generalization of the special simplifying role
played by $\xi=-1$ that corresponds to the conformally coupled scalars of Eq.(1).

In the general single-$s$ Lagrangian (\ref{action2}), field redefinitions of
the gauge invariant variable $z\rightarrow F\left(  z\right)  $ (or
$s\rightarrow\phi F\left(  s/\phi\right)  $) may be used to map one of these
three functions, $\bar{G}\left(  s/\phi\right)  ,u\left(  s/\phi\right)
,f\left(  s/\phi\right)  ,$ to any desired function of $z$ without changing
the form of the Lagrangian in (\ref{action2}). For example it may be
convenient to take $\bar{G}=1,$ and still obtain all single-$s$ Weyl-symmetric
models by using all possible $u\left(  z\right)  ,f\left(  z\right)  $.
Another option is to take all possible $\bar{G}\left(  z\right)  ,f\left(
z\right)  $ with a fixed $u\left(  z\right)  =1-z^{2}$ as in Eq.(\ref{action1}%
), which has certain advantages for understanding geodesic completeness
\cite{BCST1} of all the fields $\phi,s,g_{\mu\nu},$ in cosmological
spacetimes, as discussed in Sec. IV. If we choose $u\left(  z\right)
=1-z^{2}$ and demand renormalizability of the action in the limit of $\phi
^{2}\gg s^{2}$ (where gravity effectively decouples, as seen in the gauge
$\phi=\phi_{0}$ with $\phi_{0}\gg s$, we are forced to $\bar{G}=1$ and $f(z)$
being a quartic polynomial.

The last form is actually quite unique. We now give a proof that the general
model of Eq.(\ref{action2}) with a single $s$ (which corresponds to our
minimal realistic model) can always be written in the geodesically complete
form that we advocated in our previous work \cite{cyclicBCT}%
-\cite{cyclic-Bars} and implemented in proposing the action in
Eq.(\ref{action1}). We begin in the notation of Eq.(\ref{GenWeyl}) with only
two fields $\phi^{i}=\left(  \phi,s\right)  $ labelled by $i=0,1.$ Using well
known results of geometry in 2 dimensions, the metric in field space
$G_{ij}\left(  \phi\right)  $ can always be diagonalized by general field
reparametrizations (as in general relativity) and put into the conformally
flat form $G_{ij}=g\left(  \phi\right)  \eta_{ij}$ where $\eta_{ij}$ is the
flat metric in 2 dimensions. Using further field reparametrizations with an
overall rescaling consistent with Weyl transformations, the factor $g\left(
\phi\right)  $ can also be set equal to the constant $g\left(  \phi\right)
=1.$ Having arrived at $G_{ij}=\eta_{ij}$ as still the most general metric in
field space, the only possible form of $U\left(  \phi^{i}\right)  $ that is
consistent with the local Weyl symmetry is $U\left(  \phi^{i}\right)  =\left(
\phi^{2}-s^{2}\right)  /12.$ This completes the general proof that the action
in Eq.(\ref{action1}) is the most general Weyl invariant theory without losing
generality. Then the complete analysis of homogeneous cosmological solutions
provided in our previous work \cite{cyclicBCT}-\cite{cyclic-Bars} shows that
this form is the general geodesically complete version of the theory. Certain
other forms for $U\left(  \phi^{i}\right)  $ arrived at by field
reparametrizations, in particular purely positive forms of $U\left(  \phi
^{i}\right)  ,$ always end up putting restrictions in field space
inadvertently, and such specialized restrictions on fields is what leads to
geodesic incompleteness.

Hence, the a class of well motivated models that we used in many of our
studies amounts to $\bar{G}\left(  s/\phi\right)  =1,$ $u\left(
s/\phi\right)  =1-s^{2}/\phi^{2},$ and the low energy renormalizable quartic
potential $V\left(  \phi,s\right)  =\phi^{4}$ $f\left(  s/\phi\right)  ,$ with
$f\left(  s/\phi\right)  =\frac{\lambda}{4}\left(  s^{2}/\phi^{2}-\alpha
^{2}\right)  ^{2}+\frac{\lambda^{\prime}}{4}.$ This is the model that now is
identical to the conformally symmetric standard model we proposed. Its
homogeneous cosmological equations have been solved analytically exactly in
\cite{BCST2} for all possible initial conditions of the fields, including
radiation and curvature, and all possible values of the parameters
$\lambda,\lambda^{\prime},$ but with $\alpha=0$. After this much analytic
control, a small $\alpha$ is easily handled with numerical methods and still
have a full understanding of all the cosmological solutions of the standard
model. This forms the basis of our further work in cosmology that is discussed
in the following sections.

For further discussion, a useful gauge choice is $\phi\left(  x\right)
=\phi_{0}$ for all spacetime as in Eq.(\ref{c-gauge}). This is the gauge
called the $c$-gauge in our previous work. Because we use also other gauges,
we attach the letter $c$ to each field in this gauge, thus $\phi_{c}%
,s_{c},g_{c}^{\mu\nu}$ will remind us that we are in the $c$-gauge, where
$\phi_{c}\left(  x\right)  =\phi_{0}.$ In $c$-gauge we rename $s_{c}\left(
x\right)  =h\left(  x\right)  $ to recall that in this gauge we obtain the
simplest connection to the Higgs field $h\left(  x\right)  $ at low energy, in
nearly flat spacetime, $g_{c}^{\mu\nu}=\eta^{\mu\nu}+\cdots,$ as discussed in
the paragraphs before Eq.(\ref{unitary}). The Lagrangian in Eq.(\ref{action2})
with general $u,\bar{G},f,$ takes the following greatly simplified form in the
$c$-gauge
\begin{equation}
\mathcal{L}\left(  x\right)  =\sqrt{-g_{c}}\left[  \frac{1}{12}\phi_{0}%
^{2}u\left(  h/\phi_{0}\right)  R\left(  g_{c}\right)  -\frac{1}{2}\bar
{G}\left(  h/\phi_{0}\right)  g_{c}^{\mu\nu}\partial_{\mu}h\partial_{\nu
}h-\phi_{0}^{4}f\left(  h/\phi_{0}\right)  \right]  . \label{action3}%
\end{equation}
If we apply a Weyl transformation to go to the Einstein frame\footnote{The
Weyl transformation is $\sqrt{-g_{c}}uR\left(  g_{c}\right)  =\sqrt{-g_{E}%
}R\left(  g_{\mu\nu}^{E}\right)  +6\sqrt{u}\partial_{\mu}\left(  \sqrt{-g_{E}%
}g_{E}^{\mu\nu}\partial_{\nu}\sqrt{u}\right)  .$ After an integration by parts
of the last term (or by dropping a total derivative) we obtain the given
result.}, $g_{\mu\nu}^{E}=u\left(  h/\phi_{0}\right)  g_{\mu\nu}^{c},$ then we
obtain%
\begin{equation}
\mathcal{L}\left(  x\right)  =\sqrt{-g_{E}}\left[  \frac{1}{12}\phi_{0}%
^{2}R\left(  g_{E}\right)  -\frac{1}{2}\frac{\left(  \partial_{h/\phi_{0}%
}\sqrt{u}\right)  ^{2}+\bar{G}\left(  h/\phi_{0}\right)  }{u\left(  h/\phi
_{0}\right)  }g_{E}^{\mu\nu}\partial_{\mu}h\partial_{\nu}h-\phi_{0}^{4}%
\frac{f\left(  h/\phi_{0}\right)  }{\left(  u\left(  h/\phi_{0}\right)
\right)  ^{2}}\right]  \label{E-frame}%
\end{equation}
Without loss of generality, a particularly simplifying choice for $\bar
{G}\left(  h/\phi_{0}\right)  $, namely
\begin{equation}
\bar{G}\left(  h/\phi_{0}\right)  =u\left(  h/\phi_{0}\right)  -\left(
\partial_{h/\phi_{0}}\sqrt{u}\right)  ^{2}, \label{Gchosen}%
\end{equation}
is the one that yields a canonically normalized Higgs field. In this form, the
theory can be interpreted directly as an Einstein frame formulation of the
standard model with a canonically normalized Higgs
field\footnote{Alternatively, without fixing $\bar{G},$ it is possible to do a
field redefinition such that $h$ is written in terms of a canonically
normalized field $\sigma,$ where the relation between $\sigma$ and $h\left(
\sigma\right)  $ is given by the first order differential equation $\left(
\frac{dh}{d\sigma}\right)  ^{2}[\left(  \partial_{h/\phi_{0}}\sqrt{u}\right)
^{2}+\bar{G}\left(  h/\phi_{0}\right)  ]=u\left(  h/\phi_{0}\right)  .$ Then
the Lagrangian rewritten in terms of the canonically normalized $\sigma$ is
again of standard form with the same potential $V_{eff}\left(  h\left(
\sigma\right)  \right)  $ as Eq.(\ref{Veff}), except for expressing $h$ in
terms of $\sigma.$ This complicated procedure is avoided by the simple choice
of $G$ in Eq.(\ref{Gchosen}). \label{fieldRedef}} and an effective Higgs
potential $V_{eff}\left(  h\right)  $ given by
\begin{equation}
V_{eff}\left(  h\right)  =\phi_{0}^{4}\frac{f\left(  h/\phi_{0}\right)
}{\left(  u\left(  h/\phi_{0}\right)  \right)  ^{2}}. \label{Veff}%
\end{equation}
This potential can be crafted by various choices of $f\left(  z\right)  $ and
$u\left(  z\right)  $ to fit cosmological observations, while still being
consistent with an underlying local conformal symmetry. However, since this
gauge led to an overall positive coefficient in front of the Ricci scalar, it
could not recover all field configurations that are demanded in the
geodesically complete theory; inevitably the \textit{generic} solutions in
this gauge are geodesically incomplete as we have learned in our previous work.

For example, the best fit inflaton potentials based on recent Planck satellite
data \cite{Planck} are \textquotedblleft plateau" models\textquotedblright%
\ \cite{IjjasEtal}. Using the geodesically incomplete form of the theory in
(\ref{E-frame}) it is possible to construct examples with plateaus at large
$h/\phi_{0}$, by choosing $f$ and $u$ such that the potential $V_{eff}\left(
h\right)  $ is very slowly varying and approaches a constant at large $h$ and
by requiring $V_{eff}\left(  h\right)  $ becomes approximately the familiar
Higgs potential needed to fit low energy physics for $h\ll\phi_{0}$.

The Bezrukov-Shaposhnikov (BS) model for Higgs inflation \cite{shoposh-bez} is
a special case of our more general form in Eq.(\ref{action3}), namely their
proposal%
\begin{equation}
\mathcal{L}^{BS}\left(  x\right)  =\sqrt{-g_{c}}\left[  \frac{1}{12}\left(
\phi_{0}^{2}+\xi h^{2}\right)  R\left(  g_{c}\right)  -\frac{1}{2}g_{c}%
^{\mu\nu}\partial_{\mu}h\partial_{\nu}h-\frac{\lambda}{4}\left(  h^{2}%
-\alpha^{2}\phi_{0}^{2}\right)  ^{2}-\frac{\lambda^{\prime}}{4}\phi_{0}%
^{4}\right]  , \label{shoposh}%
\end{equation}
follows from Eq.(\ref{action3}) by taking $U\rightarrow\left(  \phi_{0}%
^{2}+\xi h^{2}\right)  ,\;\bar{G}\rightarrow1,\;$and $V\rightarrow\left(
\frac{\lambda}{4}\left(  s^{2}-\alpha^{2}\phi_{0}^{2}\right)  ^{2}%
+\frac{\lambda^{\prime}}{4}\phi_{0}^{4}\right)  .$ This leads to an effective
potential $V_{eff}$ of the type above in Eq.(\ref{Veff}),
\begin{equation}
V_{eff}\left(  \sigma\right)  =\frac{\frac{\lambda}{4}\left(  h^{2}-\alpha
^{2}\phi_{0}^{2}\right)  ^{2}+\frac{\lambda^{\prime}}{4}\phi_{0}^{4}}{\left(
1+\xi h^{2}/\phi_{0}^{2}\right)  ^{2}},\text{ with }h\rightarrow h\left(
\sigma\right)  , \label{Vshop}%
\end{equation}
In the BS model, because $G$ is chosen as $G=1$, the field $h$ is not
canonically normalized, so $h$ must be replaced through a field redefinition
\cite{shoposh-bez} to obtain a canonically normalized Higgs, $\sigma,$ as
described more generally in footnote (\ref{fieldRedef}).

Hence, the BS model has a fully Weyl symmetric formulation which was not
noticed before. Its presentation in the literature has included various
ambiguities and inconsistencies, with clashing ideas on scaling symmetries at
the classical level and quantum corrections. For example, in some cases, the
coupling to gravity has a dimensionful Newton constant that is inconsistent
with the scaling symmetry of the rest of the theory; a massless dilaton is
said to exist in cases where global scaling symmetry is explicitly broken at
low energies; unimodular gravity has been introduced but this is inconsistent
with scale symmetry; and there is ambiguity about which renormalization scheme
is appropriate for computing quantum corrections. These issues are fully
resolved with the underlying Weyl symmetric formulation discussed here. Also,
now that we have recast the BS model into a fully conformally invariant form,
we can see in the gauge fixed Einstein frame (\ref{E-frame}-\ref{Veff}) that
it is not unique. Rather, it is just a special case of a larger set of
conformally invariant models including a range that have similar plateau
properties. Furthermore, since $U=\phi^{2}+\xi h^{2}$ is purely positive, it
means the fields in the BS model form a basis that is geodesically incomplete,
and hence its generic solutions can describe only a subsector of the available
field space. We make further comments in Sec. (\ref{confCosm}) including the
effects of geodesic incompleteness.

\subsection{Supergravity \label{generalSugra}}

We do not know if supersymmetry (SUSY) is a property of nature, but
theoretically it is an attractive possibility. Therefore, it is of interest to
investigate whether it is compatible with an underlying local conformal
symmetry. As a superconformal local symmetry, the generalization of the Weyl
invariant formalism of the previous section produces stronger constraints on
scalar fields. This was derived in \cite{ibsuper} from the gauge symmetry
formalism in $4+2$ dimensions. It is possible to arrive at the results given
below directly in $3+1$ dimensions by requiring supergravity with a local
superconformal symmetry, but provided the usual Einstein-Hilbert term is
dropped (which is unusual in the supergravity literature), and instead a Weyl
symmetric formulation like the previous sections is implemented. In the $4+2$
dimensional approach of 2T-physics there is no option: it is a prediction that
the emergent $3+1$ dimensional theory is automatically invariant under a local
symmetry SU$\left(  2,2|1\right)  ,$ where SU$\left(  2,2\right)  =$SO$\left(
4,2\right)  $ is the connection to $4+2$ dimensions, the subgroup SO$\left(
1,1\right)  \subset$SO$\left(  3,1\right)  \times$SO$\left(  1,1\right)
\subset$SO$\left(  4,2\right)  $ is the Weyl subgroup that acts on the extra
$1+1$ dimensions, and the supersymmetrization in $4+2$ dimensions promotes
SU$\left(  2,2\right)  $ to SU$\left(  2,2|\mathcal{N}\right)  $ for
$\mathcal{N}$ supersymmetries \cite{2Tsusy}, with $\mathcal{N}=1$ in the
present case \cite{ibsuper}.

The scalar-field sector of the emergent $3+1$ dimensional locally
superconformal theory is presented here briefly in a streamlined fashion,
including some simplifications and extensions. First note that a scalar field
in a chiral supermultiplet must be complex. Thus SUSY requires the singlet
$\phi$ to be complexified so that it is a member of a supermultiplet. Then to
describe all the scalar fields we use a complex basis $\phi^{m}$ and denote
their complex conjugates as $\bar{\phi}^{\bar{m}}$ with the barred index
$\bar{m}.$ The notation is reminiscent of standard supergravity as reviewed in
\cite{weinberg}. However now there is no Einstein-Hilbert term; instead there
is a Weyl plus an additional $U\left(  1\right)  $ gauge symmetry that can
gauge fix a complex scalar $\phi$ into a dimensionful real constant $\phi_{0}$
that plays the role of the Planck scale.

Recalling the discussion about removing the (real) ghost scalar field $\phi$
in Sec. IIB, it should be noticed that the complex $\phi$ amounts to two real
ghosts, and therefore two gauge symmetries are needed to remove them. This
role is played by the Weyl symmetry SO$\left(  1,1\right)  $ that acts on the
extra 1+1 dimensions and the local R-symmetry U$\left(  1\right)  $, both of
which are included in the local SU$\left(  2,2|1\right)  $ outlined above. The
local U$\left(  1\right)  $ is a crucial ingredient as a partner of the Weyl
symmetry to remove the additional ghost from a complexified dilaton field
$\phi$ that is demanded by SUSY.

Like before, we have $U\left(  \phi,\bar{\phi}\right)  ,$ but SUSY requires
that the metric $G_{m\bar{n}}$ be derived from the derivatives of $U\left(
\phi,\bar{\phi}\right)  $ like a K\"{a}hler metric%
\begin{equation}
G_{m\bar{n}}=\frac{\partial^{2}U\left(  \phi,\bar{\phi}\right)  }{\partial
\phi^{m}\partial\phi^{\bar{n}}}.
\end{equation}
The metric must be non-singular and $\left(  -G_{m\bar{n}}\right)  $ cannot
contain any more than one negative eigenvalue (because no more than one
complex ghost can be removed). We denote its inverse formally as $G^{\bar{n}%
m}=\left(  \frac{\partial\bar{\phi}\otimes\partial\phi}{\partial^{2}U}\right)
^{\bar{n}m}.$ A simple quadratic example similar to Eq.(\ref{action1}) is
\begin{equation}
U\left(  \phi,\bar{\phi}\right)  =\phi^{m}\bar{\phi}^{\bar{n}}\eta_{m\bar{n}%
},\;G_{m\bar{n}}=\eta_{m\bar{n}},\text{ where }\eta_{m\bar{n}}=\text{diag}%
\left(  1,-1,\cdots,-1\right)  .
\end{equation}
In this example, all scalars in the theory are conformally coupled, and U and
G are automatically invariant under a global SU(1,N) symmetry. This global
symmetry, which will continue to be a hidden symmetry in some gauges, may be
broken explicitly by some terms in the potential $V$, depending on the model
considered. Nevertheless, keeping track of this (broken) symmetry in physical
applications can be useful.

The scalar field sector of the supergravity theory with local Weyl symmetry is
given as follows.
\begin{equation}
\mathcal{L}_{bose}=\sqrt{-g}\left[  \frac{1}{6}U\left(  \phi,\bar{\phi
}\right)  R\left(  g\right)  +\frac{\partial^{2}U}{\partial\phi^{m}%
\partial\bar{\phi}^{n}}g^{\mu\nu}D_{\mu}\phi^{m}D_{\nu}\bar{\phi}^{n}%
-V_{F+D}\left(  \phi,\bar{\phi}\right)  \right]  .
\end{equation}
The K\"{a}hler metric and K\"{a}hler potential structure is reminiscent of
general supergravity \cite{weinberg}, however the absence of the
Einstein-Hilbert term, and the corresponding scale invariance is the important
difference. Here, $U\left(  \phi,\bar{\phi}\right)  $ is homogeneous of degree
two and satisfies homothety constraints similar to Eq.(\ref{homothety}),%
\begin{equation}
U\left(  t\phi,t\bar{\phi}\right)  =t^{2}U\left(  \phi,\bar{\phi}\right)
,\;\frac{\partial U\left(  \phi,\bar{\phi}\right)  }{\partial\bar{\phi}%
^{\bar{m}}}=\phi^{n}\frac{\partial^{2}U\left(  \phi,\bar{\phi}\right)
}{\partial\phi^{n}\partial\phi^{\bar{m}}},\text{ and complex conjugate.}%
\end{equation}

The potential energy has two parts $V_{F+D}\left(  \phi,\bar{\phi}\right)
=V_{F}\left(  \phi,\bar{\phi}\right)  +V_{D}\left(  \phi,\bar{\phi}\right)  .$
The potential $V_{F}\left(  \phi,\bar{\phi}\right)  $ must be derived from an
analytic superpotential $f\left(  \phi\right)  $ that depends only on
$\phi^{m}$ and not on $\bar{\phi}^{\bar{m}}$, and is homogeneous of degree
three so that $V_{F}$ is homogeneous of degree four
\begin{equation}
f\left(  t\phi\right)  =t^{3}f\left(  \phi\right)  ,\;\;V_{F}=-\left(
\frac{\partial\bar{\phi}\otimes\partial\phi}{\partial^{2}U}\right)  ^{\bar
{n}m}\frac{\partial\bar{f}}{\partial\bar{\phi}^{\bar{n}}}\frac{\partial
f}{\partial\phi^{m}}.
\end{equation}
The potential $V_{D}\left(  \phi,\bar{\phi}\right)  $ is derived from another
independent analytic function $z_{ab}\left(  \phi\right)  $ with adjoint group
indices $a,b$ that appears in Yang-Mills terms as, $-\frac{1}{4}%
\operatorname{Re}\left(  z_{ab}\left(  \phi\right)  \right)  F_{\mu\nu}%
^{a}F^{b\mu\nu}$ . This $z_{ab}\left(  \phi\right)  $ should be homogeneous of
degree zero $z_{ab}\left(  t\phi\right)  =z_{ab}\left(  \phi\right)  .$ Then
$V_{D}$ takes the form
\begin{equation}
V_{D}\left(  \phi,\bar{\phi}\right)  =\frac{1}{2}\operatorname{Re}\left(
z_{ab}^{-1}\right)  \left(  \frac{\partial U}{\partial\phi^{m}}\left(
t_{a}\phi\right)  ^{m}\right)  \left(  \frac{\partial U}{\partial\bar{\phi
}^{\bar{n}}}\left(  t_{b}\bar{\phi}\right)  ^{\bar{n}}\right)  ,
\end{equation}
where $t_{a}$ is the matrix representation of the Yang-Mills group as it acts
on the scalars $\left(  \phi^{m},\bar{\phi}^{\bar{m}}\right)  .$ The fermionic
terms are added consistently with the usual rules of supergravity
\cite{weinberg}.

This is the general setup for the Weyl invariant matter coupled to Weyl
invariant supergravity as derived from the $4+2$ dimensional theory. The
simplest example that corresponds to the supersymmetric generalization of
Eq.(\ref{action1}) would be the minimal supersymmetric standard model extended
with an additional singlet supermultiplet, whose scalar component is the
complex field $\phi.$ Then $U$ takes the quadratic form suggested in
\cite{ibsuper}
\begin{equation}
U=\bar{\phi}\phi-H_{u}^{\dagger}H_{u}-H_{d}^{\dagger}H_{d},
\end{equation}
where $H_{u},H_{d}$ are the two Higgs doublets needed in the supersymmetric
version of the standard model. The superpotential $f\left(  \phi,H_{d}%
,H_{u}\right)  $ and the matrix $z_{ab}\left(  \phi,H_{d},H_{u}\right)  $ are
chosen to fit the usual practice of supersymmetric model building
\cite{weinberg}. An example is $z_{ab}=\delta_{ab}$, and $f\left(  \phi
,H_{d},H_{u}\right)  =g\left(  H_{u}H_{d}\right)  \phi+g^{\prime3},$ where
$\left(  H_{u}H_{d}\right)  =H_{u}^{\alpha}H_{d}^{\beta}\varepsilon
_{\alpha\beta}$ is the only SU$\left(  2\right)  \times$U$\left(  1\right)  $
invariant which is analytic in both $H_{u}$ and $H_{d}$. In an effective
(rather than renormalizable) low energy theory, these can be modified by
replacing the dimensionless coupling constants $g,g^{\prime}$ by an arbitrary
function of the ratio $\left(  H_{u}H_{d}\right)  /\phi^{2},$ and similarly
for $z_{ab}.$ When the complex $\phi\left(  x\right)  $ is gauge fixed to the
real constant $\phi_{0},$ this approach generates all the dimensionful
parameters from the same source $\phi_{0}$ which is of order of the Planck
scale. Then we see that the modification of $g,g^{\prime}$ and $z_{ab}$ by
arbitrary functions of $\left(  H_{u}H_{d}\right)  /\phi^{2}$ is negligible at
low energies.

More general scale invariant models with more fields (such as generalizations
of the minimal SUSY model) are easily constructed by using the rules on scalar
fields given in this section. With special forms of the superpotential
$f\left(  \phi\right)  $, it is possible to construct so called
\textquotedblleft no-scale\textquotedblright\ models \cite{noScale1}%
-\cite{noScale3} in which the cosmological constant is guaranteed to be zero
at the classical level even after spontaneous breakdown of symmetries. Simple
examples of no-scale models, that are lifted to be fully Weyl invariant, are
given in \cite{ibsuper}.

In this paper we will not explore any further the general superconformal
structures discussed above although this could be of interest in some future
applications. Some of our examples that are very similar to the Weyl invariant
supersymmetric cases previously discovered in \cite{ibsuper} were later
explored in a cosmological context in \cite{feraraLindeKal}. However their
discussion, which is focussed on very specialized initial conditions that are
so non-generic, is difficult to be convincing especially in the face of the
complete set of solutions that we now understand much better.

\section{Conformal Cosmology \label{confCosm}}

In this paper we have emphasized the geodesically complete nature of our
conformal standard model. In this section we first describe what we mean by
geodesic completeness and the related notion of completeness in field space,
and then discuss some applications of our conformal standard model in cosmology.

There are two required properties for geodesic completeness: the first is
smooth local geodesic continuation through singularities across all space-time
patches; the second is infinite action for geodesics that reach arbitrarily
far into the past so that unnatural initial conditions are avoided. Both
properties are satisfied in our theory which is insured to also be complete in
field space. Geodesics $x^{\mu}\left(  \lambda\right)  $ are computed by
extremizing the particle action $S=-\int d\lambda~m\sqrt{\dot{x}^{\mu}\dot
{x}^{\nu}g_{\mu\nu}\left(  x\left(  \lambda\right)  \right)  }$ where $m$ is
the mass and $g_{\mu\nu}$ is a gravitational field. In our Weyl invariant
standard model (\ref{action1}) all particle masses are generated by the Higgs
field, so $m$ is proportional to the Higgs field $h$, $m=gh\left(  x\left(
\lambda\right)  \right)  ,$ where $g$ is the dimensionless coupling constant
for the corresponding particle as prescribed by the standard model. Then this
particle action is locally invariant under Weyl transformations $h\left(
x\right)  \rightarrow\Omega\left(  x\right)  h\left(  x\right)  $ and
$g_{\mu\nu}\left(  x\right)  \rightarrow\Omega^{-2}\left(  x\right)  g_{\mu
\nu}\left(  x\right)  .$ For our cosmological discussion, geodesics $x^{\mu
}\left(  \lambda\right)  $ in a spatially homogeneous space are computed when
both homogeneous fields that appear in the particle action $S$, $h\left(
\tau\right)  $ and $g_{\mu\nu}\left(  \tau\right)  =a^{2}\left(  \tau\right)
\left(  \eta_{\mu\nu}+\text{anisotropy \& curvature}\right)  ,$ where
$\tau\equiv x^{0}$ is the conformal time, correspond to consistent solutions
of the field equations of our conformal standard model in (\ref{action1}). The
full solution to the geodesic equation (neglecting curvature and anisotropy
for simplicity) is \
\begin{equation}
\vec{x}\left(  \tau\right)  =\vec{q}+\vec{p}\int_{\tau_{0}}^{\tau}\frac
{d\tau^{\prime}}{\sqrt{\vec{p}^{2}+m^{2}\left(  \tau^{\prime}\right)
a^{2}\left(  \tau^{\prime}\right)  }}.\label{geodesicX}%
\end{equation}
where $\vec{p}$ is the conserved spatial component of the particle momentum
vector, $p^{\mu}=\partial S/\partial\dot{x}^{\mu},$ while $p^{0}=\sqrt{\vec
{p}^{2}+m^{2}\left(  \tau\right)  a^{2}\left(  \tau\right)  }$, and $\vec
{q}=\vec{x}\left(  \tau_{0}\right)  $ is the initial position. These
expressions are invariant under reparametrizations of the affine parameter
$\lambda,$ since they depend only on the physical spacetime variables $x^{\mu
}.$ Then, the local continuation property of a geodesic $\vec{x}\left(
\tau\right)  $ in a homogeneous space is satisfied automatically when one
insures that all the fields, including $m\left(  \tau\right)  =gh\left(
\tau\right)  $ and $a\left(  \tau\right)  ,$ are given in all the patches of
homogeneous field space and that the fields are smoothly continued through
singularities. In \cite{BCST1}\cite{BCST2}, we showed how local Weyl
invariance, in the special Weyl frame displayed in Eq.(\ref{action1}) that
covers all the patches, are essential ingredients to insure that all
homogeneous cosmological solutions of our conformal standard model
(\ref{action1}) have this continuation property.

The global action for a geodesic (\ref{geodesicX}) is given by
\begin{equation}
\left\vert S_{fi}\right\vert =\int_{\tau_{i}}^{\tau_{f}}d\tau\frac
{m^{2}\left(  \tau\right)  a^{2}\left(  \tau\right)  ~}{\sqrt{\vec{p}%
^{2}+m^{2}\left(  \tau\right)  a^{2}\left(  \tau\right)  }}.\label{geodTime}%
\end{equation}
The combination $m^{2}\left(  \tau\right)  a^{2}\left(  \tau\right)
=g^{2}h^{2}\left(  \tau\right)  a^{2}\left(  \tau\right)  $ that appears in
$\left\vert S_{fi}\right\vert $ is invariant under Weyl transformations, so it
can be computed in any convenient gauge (see e.g. Eqs.(\ref{EgaugeU}%
-\ref{detxU})) . In particular in the $\gamma$-gauge used in the solutions in
\cite{BCST1}\cite{BCST2}, where $a_{\gamma}\left(  \tau\right)  =1,$ we have
shown that in all cyclic solutions in \cite{BCST2}, $h^{2}\left(  \tau\right)
a^{2}\left(  \tau\right)  =h_{\gamma}^{2}\left(  \tau\right)  a_{\gamma}%
^{2}\left(  \tau\right)  $ oscillates with a fixed maximum amplitude many
times within a cycle, and for an infinite number of times over an infinite
number of cycles, leading to an infinite action, since $\left\vert
S_{fi}\right\vert $ receives the same positive finite contribution in each
cycle. For these solutions the average magnitude of
$\vert$%
$a_{E}$%
$\vert$
in the $E$-gauge (usual Einstein frame, see Eq.(\ref{EgaugeU})) is the same in
each cycle. Recently, we have also constructed \cite{BST-HiggsCosmo} solutions
with increasing entropy in each future cycle such that the average
$\vert$%
$a_{E}$%
$\vert$
in the $E$-gauge increases in each future cycle. For these solutions,
$\left\vert S_{fi}\right\vert $ diverges to the past \cite{BST-HiggsCosmo}
even faster as compared to our solutions in \cite{BCST2}. Hence, unlike
inflation scenarios \cite{Borde:2001nh}, we may expect that the universe
described by our conformal standard model has no unnatural initial conditions.

Next, we discuss various potential applications of Weyl invariant theories to
cosmology. Early universe cosmology near the singularity is a natural place to
look for uses because the interactions imposed by Weyl invariance (more
profoundly $4+2$ dimensional gauge symmetries) produce their most significant
changes relative to conventional physics in the limit of strong gravity, as
noted in some of the examples above. Black hole phenomena is another
interesting arena for study, but we will not discuss them here.

The first important application was to use Weyl invariance to construct
geodesically complete cosmological solutions. Conventional cosmological
analysis is usually confined to theories coupled to Einstein gravity with a
dimensionful Newton's constant. In this framework, all cosmological solutions
of interest are geodesically incomplete. However, we have observed that a
Weyl-invariant theory can be cast in Einstein frame, as illustrated by
Eq.(\ref{E-frame}), and in other frames. The observation made in
\cite{cyclicBCT}-\cite{BibBang-IB} is that geodesically incompleteness is an
artifact of an unsuitable frame choice: geodesically incomplete solutions in
Einstein frame may be completed in other frames, even though the theories are
entirely equivalent away from the singularity.

The Einstein frame can always be reached directly from a Weyl invariant theory
in (\ref{action2}) by making the Einstein-gauge choice
\begin{equation}
\frac{1}{12}U\left(  \phi_{E},s_{E}\right)  =\left(  16\pi G\right)
^{-1}=\frac{1}{2}. \label{EgaugeU}%
\end{equation}
We label this the $E$-gauge and mark the fields in this gauge with the
subscript $E$ (as in $\left(  \phi_{E},s_{E},g_{E}^{\mu\nu}\right)  $) to
distinguish them from the $c$- gauge fields $\left(  \phi_{c},s_{c},g_{c}%
^{\mu\nu}\right)  .$ Note that the $E$-gauge can be valid only in a patch in
field space since restrictions must be imposed on the fields to require that
$U\left(  \phi_{E},s_{E}\right)  $ is positive.

The same theory can easily be transformed to other gauges. For example, the
relations between the $E$- and $c$-gauge fields can be easily derived by
considering the Weyl gauge invariants such as $s/\phi$ and $\left(
\det\left(  -g\right)  \right)  ^{1/8}\phi,$ $\left(  \det\left(  -g\right)
\right)  ^{1/8}s,$ etc., for example we deduce
\begin{equation}
\frac{s_{E}}{\phi_{E}}=\frac{h}{\phi_{0}},\;\text{and }\left(  \det\left(
-g_{c}\right)  \right)  ^{1/8}\phi_{0}=\left(  \det\left(  -g_{E}\right)
\right)  ^{1/8}\phi_{E},\text{ etc.}%
\end{equation}
From these, we can express $\phi_{E}\left(  h\right)  $ and $s_{E}\left(
h\right)  $ in terms of the single field $h,$ so that the gauge condition
(\ref{EgaugeU}) is satisfied. Inserting these expressions into the gauge
invariant action (\ref{action2}) we arrive at the same $E$-frame action as
Eq.(\ref{E-frame}).

As argued in our work \cite{cyclicBCT}-\cite{BibBang-IB}, classically
geodesically complete solutions can be obtained for all single-scalar theories
that can be cast in the form of Eq.(\ref{action2}). But, for all patches of
field space $\left(  \phi,s,g_{\mu\nu}\right)  $ to be included, as demanded
by the geodesics derived from $V_{eff}\left(  h\right)  $ in the Einstein
frame, $\left(  U,G,f\right)  $ must be brought to an appropriate form by
using the field redefinitions discussed below Eq.(\ref{xi}). The patches of
field space $\left(  \phi,s,g_{\mu\nu}\right)  $ that are missing in the
Einstein frame can then be added in order to obtain a geodesically complete
space. We have argued in \cite{BCST1}, and provided a proof in section
(\ref{generalGrav}), that geodesic completeness is accomplished when we bring
$U$ to the form $U=\phi^{2}-s^{2},$ where $s^{2}=\sum s_{I}^{2},$ is the sum
of all scalars other than $\phi.$

Geodesics remain incomplete when they hit the singularity in all frames in
which $U$ is always positive, or the equivalent Einstein frame, such as the
one in Eq.(\ref{shoposh}). When we rewrite those, by field redefinitions, in
terms of new fields in which $U=\phi^{2}-s^{2},$ then the patches of field
space $\phi^{2}\geq s^{2}$ are equivalent to the Einstein frame, or to the
other frames with positive $U$. Geodesic completion is achieved by allowing
all regions of field space in the parametrization that has $U=\phi^{2}-s^{2};$
for this form, $U$ is allowed to smoothly go negative. In these coordinates
the gauge invariant vanishing point of $U=0,$ given by the gauge invariant
expression $\left\vert s/\phi\right\vert =1,$ has a special significance; it
corresponds to the singularity of the scale factor of the universe in the
Einstein gauge. This is seen by equating the gauge invariant $\left(
\det\left(  g\right)  \right)  ^{1/4}U\left(  \phi,s\right)  $ in the Einstein
gauge, in which $U\left(  \phi_{E},s_{E}\right)  =6,$ and the unimodular gauge
(labelled with $\gamma$), in which $\det\left(  -g_{\gamma}\right)  =1,$ as
follows
\begin{equation}
\left(  \det\left(  -g\right)  \right)  ^{1/4}U\left(  \phi,s\right)  =\left(
\det\left(  -g_{E}\right)  \right)  ^{1/4}\times6=1\times U\left(
\phi_{\gamma},s_{\gamma}\right)  .\label{detxU}%
\end{equation}
At spacetime points or regions where $U\left(  \phi_{\gamma},s_{\gamma
}\right)  =\phi_{\gamma}^{2}-s_{\gamma}^{2}=0,$ which is where the gauge
invariant quantity $\left\vert s/\phi\right\vert $ hits unity in all gauges,
$\left\vert s_{\gamma}/\phi_{\gamma}\right\vert =$ $\left\vert s_{E}/\phi
_{E}\right\vert =\left\vert h/\phi_{0}\right\vert =1,$ the geometry in the
Einstein gauge fails completely since $\det\left(  -g_{E}\right)  =0$, and
this is how the cosmological singularity occurs at some point in time
\cite{BCST1}. Then, as we can see in the $c$-gauge, the region $h/\phi_{0}%
\sim1$ (Higgs of Planck size) is the region of the big crunch or big bang
singularity, where $\left\vert s/\phi\right\vert =1$ in any gauge. The
behavior of the universe in this region is governed by a universal attractor
mechanism that is independent of the scalar potential, therefore independent
of the details of the model \cite{BCST1}.

Although in the classical theory, cosmological singularities of FRW-type are
typically resolved in the Weyl-lifted theory with $U=\phi^{2}-s^{2}$ in
suitable Weyl gauges, one should still worry about quantum gravity
corrections. When $U$ vanishes, the coefficient of the Ricci scalar in the
gravitational action vanishes so there is no suppression of metric
fluctuations and quantum gravity corrections should become large. However, it
is notable that for certain types of cosmic singularities, including realistic
ones, the metric and fields possess a unique continuation around the
singularity in the complex time plane. A complex time path can be chosen to
remain far from the singularity so that $U$ is always large and gravity
remains weak. Thus, we are able to find an analytic continuation of our
classical solutions connecting big crunches to big bangs along which quantum
gravity corrections are small. This issue is under active investigation and we
defer further discussion to future work.

A particularly interesting application of Weyl-invariance is to Higgs
cosmology, which the geodesically complete solutions show can be much more
interesting than conventionally assumed. In fact, there is the possibility
that the Higgs field alone may be sufficient to explain the large-scale
features of the universe, as suggested by the Bezrukov-Shaposhnikov Higgs
inflation model \cite{shoposh-bez} and our recent work on the Higgs in cyclic
cosmology \cite{lectures}\cite{BST-HiggsCosmo}.

In considering Higgs cosmology or other models of the early universe, we
believe that the cosmological solutions found for geodesically complete
theories provide some important new insights on the question of the likely
initial conditions just after the big bang. For example, the \textit{generic}
behavior of the Higgs after the bang when $h\sim\phi_{0}$ is dramatically
different than the contrived initial conditions that are commonly assumed in
Higgs inflation scenarios. This is easily seen by examining the analytic work
in \cite{BCST1}\cite{BCST2}, where for the simple model in Eq.(\ref{action1})
with $\alpha=0,$ all homogeneous solutions including curvature and radiation
are obtained and the effects of anisotropy are determined \cite{BCST1}.
Emphasizing that this is not only the contrived solutions, but all solutions,
it serves as an example of the richness of the phenomena that occur in a model
even with a simple potential. The results make clear a point that is obvious,
but often forgotten: for a given scalar potential, there is an enormous range
of cosmological solutions. By comparison, it is clear that the slow-roll
initial conditions frequently assumed in the analysis of cosmic inflation are
very special and unlikely. For example, by transforming the inflationary
solution in the Bezrukov-Shaposhnikov Higgs inflation model \cite{shoposh-bez}
to a field basis in which $U=\phi^{2}-s^{2}$, it may be possible to trace
cosmic evolution right back to the singularity and to judge whether the
inflation is likely in the space of geodesically complete cosmological solutions.

Fig.(1) is an illustration of our solution for the \textit{generic}
cosmological behavior of the Higgs field just after the big bang if the Higgs
potential has a stable non-trivial minimum, as in Eq.(\ref{action1}), as
usually assumed and as required for the Bezrukov-Shaposhnikov Higgs inflation
model \cite{shoposh-bez}. This figure describes the generic cosmological
evolution of the Higgs field, that must start with fluctuations of Planck size
and energy (due to the universal attractor near the singularity \cite{BCST1}),
and quickly reduce its amplitude by losing energy to the gravitational field;
then after a phase transition, settle down to an almost constant value at the
electroweak scale $v$ determined by the dimensionless parameter $\alpha$ in
Eq.(\ref{alpha}).%

\begin{center}
\includegraphics[
height=1.9121in,
width=3.2214in
]%
{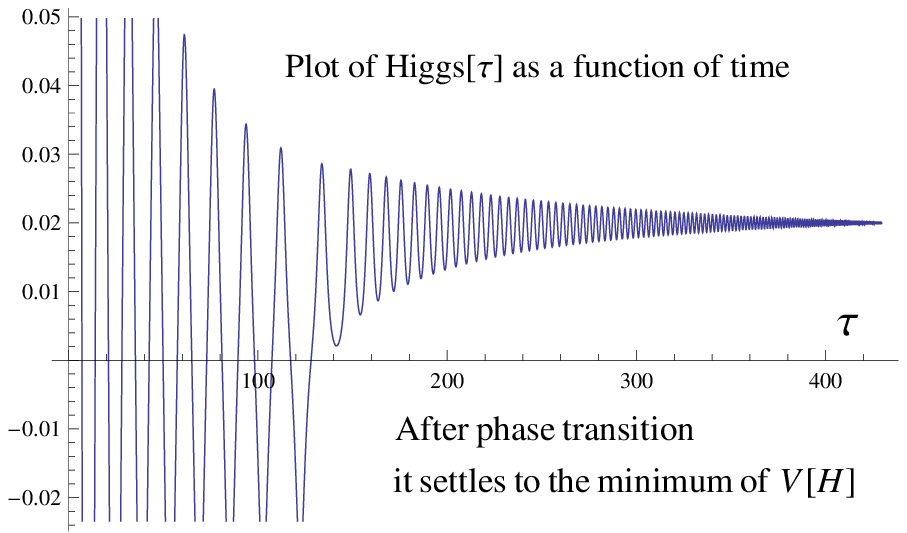}%
\\
Fig.(1) - After the Big Bang the Higgs field oscillates initially around zero
with a large amplitude of the order of the Planck scale. It slowly loses
energy to the gravitational field, causing its amplitude to diminish. As it
approaches the time or energies of the electroweak scale, it undergoes the
phase transition seen in the figure, and then slowly settles to a constant
vacuum value $v$ at a stable minimum of the potential.
\label{cosmohiggs-stable.eps}%
\end{center}

The solution of Fig.(1) changes drastically if the vacuum is metastable after
including quantum corrections, which is a possibility suggested by the most
recent LHC data for the Higgs and the top quark masses \cite{instabilityHiggs}%
, and assuming no new physics up to the Planck scale. Metastability is
incompatible with Higgs inflation and generally causes problems for inflation
because the Higgs will typically escape from the metastable phase right after
the big bang and roll to a state of negative energy density that can prevent
inflation of any sort from occurring.

The exact solutions of the Weyl-invariant theory suggest an alternative
cosmology in this case. The generic solution at first behaves as in Fig.(1)
after the big bang, all the way through the electroweak phase transition. But
after some time (order of the lifetime of the universe) the Higgs oscillations
in the electroweak vacuum grow larger and larger, like the mirror image of
Fig.(1), taking away energy from the gravitational field and eventually
causing a collapse of the universe to a big crunch, while the Higgs does a
quantum tunneling to a lower state of the potential. At that stage our exact
analysis near the singularity given in \cite{BCST1} takes over to describe
interesting new phenomena that occur just after the crunch and before another
rebirth of the universe with a big bang. The result is a regularly repeating
sequence in which the Higgs field is trapped in its metastable state after a
big bang, remains there for a long period of expansion followed by
contraction, escapes as the universe approaches the big crunch, passes through
to a big bang and becomes trapped again. The evolution can be considered a
Higgs-driven cyclic theory of the universe. The details are presented in a
separate paper \cite{BST-HiggsCosmo}.

\begin{acknowledgments}
This research was partially supported by the U.S. Department of Energy under
grant number DE-FG03-84ER40168 (IB) and under grant number DE-FG02- 91ER40671
(PJS). Research at Perimeter Institute is supported by the Government of
Canada through Industry Canada and by the Province of Ontario through the
Ministry of Research and Innovation. The work of NT is supported in part by a
grant from the John Templeton Foundation. The opinions expressed in this
publication are those of the authors and do not necessarily reflect the views
of the John Templeton Foundation. IB thanks CERN and PJS thanks Perimeter
Institute for hospitality while this research was completed.
\end{acknowledgments}

\end{document}